# Design and Fabrication of Novel Digital Transcranial Electrical Stimulator for Medical and Psychiatry Applications

HoseinAli Jafari[1], Mohammad Bagher Heydari[1,*], Niloofar Jafari[2], Hamid Mirhosseini[3]

[1,*] School of Electrical Engineering, Iran University of Science and Technology, Tehran, Iran
[2] School of Advanced Medical Sciences, Tabriz University of Medical Sciences, Tabriz, Iran
[3] Research Center of Addiction and Behavioral Sciences, Shahid Sadoughi University of Medical Sciences, Yazd, Iran

[*] E-mail: mo_heydari@elec.iust.ac.ir

*Abstract—* **In this article, we design a novel Transcranial Electrical Stimulator for medical applications, which is very cheap and can produce the desired signals very accurately. Our fabricated stimulator generates all current signals related to Transcranial Electrical Stimulation (TES) methods, i.e. Transcranial Direct Current Stimulation (tDCS), Transcranial Pulsed Current Stimulation (tPCS), Cranial Electrotherapy Stimulation (CES), and Micro-current Electrical Therapy (MET). The proposed stimulator has been constructed of an advanced digital controller which makes it tunable. One of the major advantages of the device is its ability to generate Burst pulses.**

## I. Introduction

In recent years, electrical interventions for cognitive rehabilitation have been vastly developed. These methods are fairly chip and can be applied easily on nearly everyone except for some special groups like pregnant women. Using electrical current flow for diagnosing and therapy in medicine has a long time background. In 1755, Charles le Roy, a French doctor, wound cords with battery around the head of a man to cure the vision of his blind patient. But in the 19$^{th}$ and 20$^{th}$ centuries, studies about the healing of psychiatric disorders with electric current greatly increased, so Shock-therapy emerged in the 1930's [1-3]. The first studies on patients with strong electric current were held in 1939 for epileptogenic activities [4].

Nowadays, many different techniques are suggested for investigating mental disorders [5-7]. Using low-intensity current for checking and healing of psychosis firstly introduced by Nitsche and Paulus in 2000 [8]. Overall, there are four methods for transcranial electrical stimulation with low intensity: Direct Current, Alternating Current, Pulsed Current, and Random noise current transcranial stimulation.

The usage of direct current (DC) signals returns to the early 1800s when Giovanni Aldini published some reports about curing of depression. In 2000, Nitsche and Paulus showed that transcranial stimulation by direct current has no adverse effects on the brain [8]. Afterward, computer-simulated models proved that this kind of stimulation may induce significant current flows in surface layers of the brain cortex [9-14]. Recently, by discovering benefits of this method, many research articles were published in this field [15]. At first steps, there were some limitations for focus because of the large area pads. Lately, to improve the concentration, a set of High Definition electrodes with multiple output were used [16]. This method has many applications nowadays in psychiatry such as post-stroke motor rehabilitation [17], improving behavioral performance in Alzheimer's disease [18], alleviation of chronic pain [19].

The other transcranial stimulation is the usage of alternating current. This way can change the brain's single neuron's transit potential. For instance, a sine/cosine wave current may induce an oscillation in the brain's cortex potential [20]. Pulsed current is utilized for stimulating the brain's cortex and sub-cortical layer [21]. Similar to the direct current, various applications can be gained by positioning the electrodes on different points of the head in this method. Cranial electrical stimulation (CES) is the well-known form of the pulsed current stimulating with several applications. Usage of the random noise current signal for transcranial stimulation was firstly introduced by Terney [22]. He showed that the brain's cortex can be much more stimulated with a current signal of random frequency and domain with custom intensity [23].

In this paper, a new device will be introduced that can produce four different types of signals for transcranial stimulation. Output signals are of current-mode [24-26] but instead from a simple and reliable voltage-mode circuit controlled by two digital processors yield to highly safe operating and very low production cost in contrast with its commercial counterparts. Based on achieved results from experimental use, very accurate signals with less than 3% tolerance in timing and intensity and high tunability is eminent that makes it very desirable for therapists and more the researchers. Long-lasting battery charge for more than a full day that came true by low power consumption, and having several different uses in a single device make it very suitable for clinical use. Providing an electric current signal instead of voltage, and tuning voltage for up to 30V automatically, make it governable and safe the most. In other words, there is not any voltage supply greater than 30V, and the current signal is made through it, and dropped voltage across the body is applied by physic laws and always remains under 30V. In the first section, a brief history of the electrical brain stimulation is reviewed. Four well-known technics of transcranial stimulation is theoretically described in section two with their implemented circuit considerations. Section three shows the results achieved by utilizing the proposed device on a resistive load. Finally, the main points and achievements in contrast to some devices in the market of IRAN are briefly discussed.

## II. FUNCTIONING MODES

The proposed device uses two digital processors as master and slave for better controlling over output signal parameters such as timing and intensity. The timing signals provided by the slave processor, control output current flow switches by turning them ON or OFF.

Intensity is controlled by a simple DAC consists of an RC low pass filter (LPF) that averages PWM signal from 0 to 100% into a DC voltage level of GND to VCC. The provided DC voltage level will be interpreted through a simple voltage-to-current converter (Fig.1) composed of a common-emitter NPN transistor with a resistor in its Emitter branch that defines the trans conductance proportion as follows:

$$I_{out} = \frac{V_{intensity} - v_{BE(ON)}}{R_E} \quad (1)$$

In addition to the VCC supply for digital circuits and processors, there is a 30V power supply that empowers the output stage for high resistive load and large enough currents up to the safety consideration limits. The 30V supply itself is provided by the usage of a low power SMPS DC-DC voltage booster that has a good efficiency with small size and low cost with no risk of any danger or unsafe.

While the applied current intensity is flowing through the collector of a common-emitter NPN transistor (Fig. 1) and 30V supply, the dropped voltage across the body will be less than 30V under any condition and can be calculated:

$$V_{body} = I_{out} \times R_{body} \quad (2)$$

Two extra NPN transistors are added (Fig. 2) for better accuracy and faster control over the applied current with the cost of available voltage decrement and added error calculated in (3) and (4), respectively. Since the LPF output voltage is a slow signal, that is not possible to make a pulsed current with changing of Pulse width Modulation(PWM) signal but it is done through a control signal that goes T1 from the cut-off region into saturation and allows the output current to flow and vice versa.

$$v_{AVAILABLE} = 30V - (v_{CC} - v_{BE(ON)T3} - v_{CE(sat)T3}) \quad (3)$$

$$I_{out} = I_{out(target)} - \frac{v_{CE(sat)T1}}{R_E} \quad (4)$$

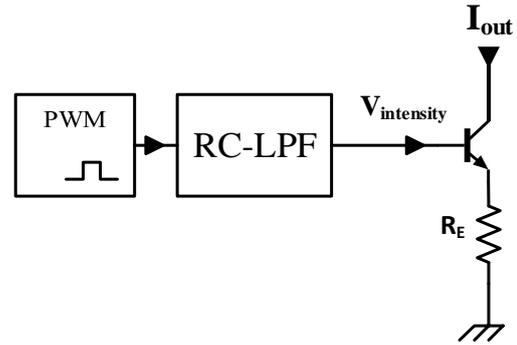

**Fig. 1.** Voltage to current converter

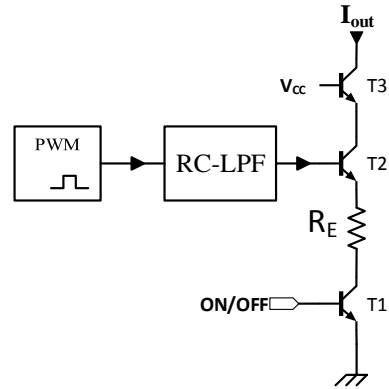

**Fig. 2.** The controllable buffered current output stage

Transistor T3 fixes the collector voltage of T2 to $V_{CC}-V_{BE(ON)}$ which is almost constant and suppresses $V_{CE}$ variation on T2 that yields to independent current $I_{out}$ (4) from load resistance and voltage drop:

$$I_{out} = \left(I_{out(target)} - \frac{v_{CE(sat)T1}}{R_E}\right).\left(1 + \frac{(v_{CC}-v_{BE(ON)T3})-(v_{intensity}-v_{BE(ON)T2})}{V_A}\right) \quad (5)$$

The intensity schedule and waveform of the output signal for four different types of therapies which is implemented in the proposed device are described as follows.

*A. tDCS*

Direct current stimulation has no frequency at all and the applied signal is always flowing through the body from anode to cathode and never goes in a reverse direction. For some medical reasons, the applied current would increase gradually from zero to a determined level by a rate of 1mA per minute as sketched in Fig. 3. Rising ($T_W$) and falling ($T_C$) slopes are called "warming up" and "cooling down" respectively. The prescribed dose of current for healing [10-11] will be applied during $T_D$. This procedure is also applied to other methods of intensity control. By enabling SHAM mode, the intensity during the TD will turn to zero silently without any sign to be able to investigate the placebo effect in research proposes.

*B. tPCS*

Pulsed current stimulation in addition to the intensity that changes in a form like tDCS, has a frequency and Duty Cycle range which expands in 0.5Hz to 1000Hz and 10% to 90% respectively. Duty-cycle (D.C.) will remain constant during the process ($T_D$) but the frequency is varying randomly between the lower and higher limits those are defined by the user. The signal parameters for a single pulse of tPCS is shown in Fig.4 and calculated in (5) and (6) that $T_P$ is the length of a single pulse period:

$$DutyCycle_\% = \frac{T_{ON}}{T_{ON}+T_{OFF}} \times 100 \quad (6)$$

$$Frequency = \frac{1}{T_{ON}+T_{OFF}} = 1/T_P \quad (7)$$

*C. CES*

CES pulses are alternating repeatedly in two directions as for left and right pulses that are implemented using an H-bridge with a tailed current source as intensity controller. H-bridge is composed of 2 NMOS and 2 PMOS for its switching elements. The produced signal can also be patterned in addition to set intensity, frequency range, and duty cycle. Feasible patterns are Random, Frequency modulation, and burst, all are drawn conceptually in Fig. 5 with contrast to a non-patterned single frequency pulse called "continuous". In frequency modulation (F.M.) pattern, each pulse frequency or period, calculated in a regular manner that moves from the lowest value to the top one in the defined range gradually and then returns to the start value and this will keep going on.

As it is obvious in Fig. 5 the burst pattern has two extra parameters that have to be determined, burst frequency ($f_B$) and pulse chain counts (N). In this pattern, the basic pulses are repeated normally for N times that must be greater than unity. After that, the output signal stops until the end of the burst period. The burst period should be at least more than two times greater than the maximum period of the basic pulses.

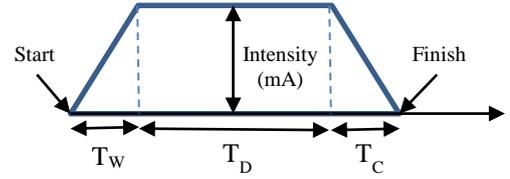

Fig. 3. tDCS current amplitude over time

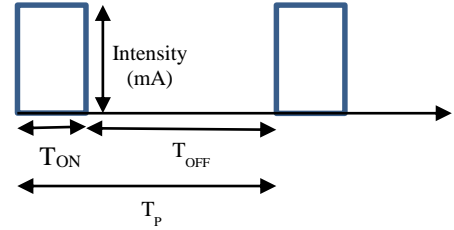

Fig. 4. tPCS signle pulse parameters

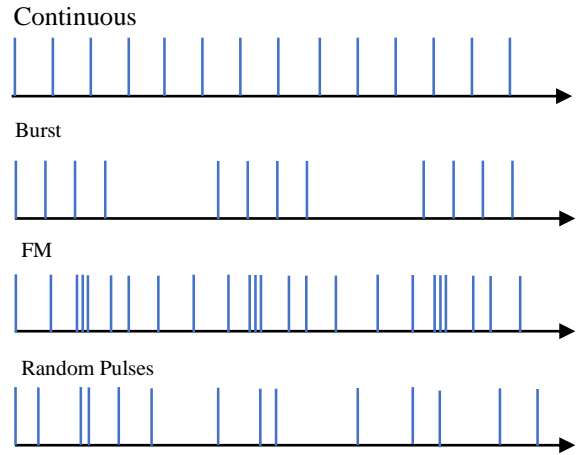

Fig. 5. Patterns in CES mode

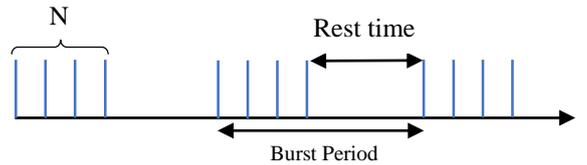

Fig. 6. Burst pattern parameters

*D. MET*

This method works the same as CES but with fixed predefined parameters for frequency range, pattern, and pulse width. Because of its very narrow pulses, the average intensity applied is in the range of microampere to affect the cells rather than CES that normally works in the mA range.

## III. SIMULATION AND MEASUREMENTS

The proposed device is designed and simulated using Proteus version 7.10 ISIS/ARES for schematic and its printed circuit board (PCB) as shown in Fig 7.

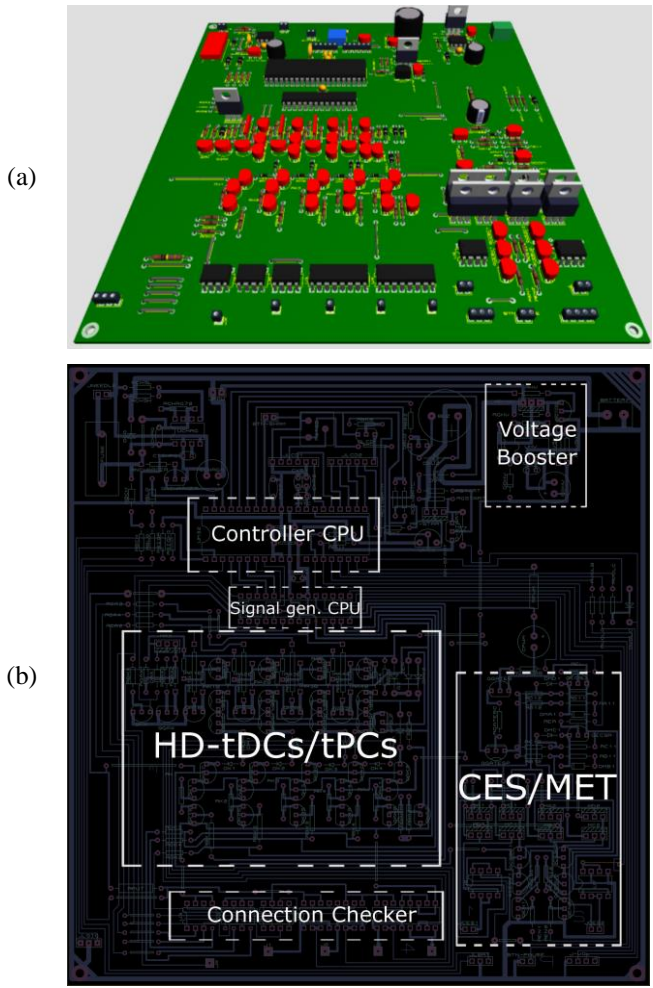

**Fig. 7.** printed circuit board designed by proteus ARES in 18cm×19cm (a) 3D visualisation and (b) partly described

Measurements are also done by the use of an oscilloscope (Fig. 8) for CES output signals. There is a very good match between expected and achieved results. Output current flows through a 10KΩ instead of a real human body. The intensity level is tunable in the range of 0.1 to 4.0 mA by 0.1mA steps. The frequency range of the output signals is limited from 0.5Hz to 1000Hz for basic pulses and from 1Hz to 20Hz for burst frequency in that pattern with customizable pulse train chains up to 15.

Random noise signal passed through a high cut-off LPF to guarantee the diminishing of high-frequency components and also programmed to limit its bandwidth not more than 300Hz. Fast Fourier Transform (FFT) on produced noise signal shows (Fig. 8) that most signal energy is concentrated below 300Hz.

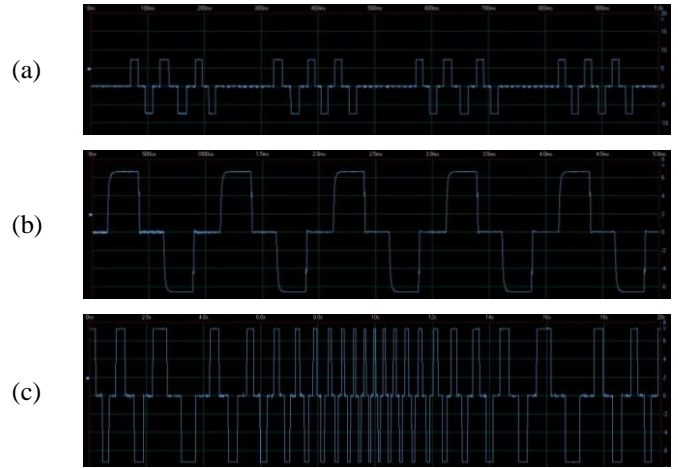

**Fig. 8.** Voltage droped on output load connections in (a)burst, (b)single frequency and (c)F.M. pattern

## IV. CONCLUSION

In this paper, a well-designed device having high tunable options with reliable and safe output is proposed and fabricated in a very simple manner with general propose elements to lower cost and revive the supply chain. Frequency, intensity, and duty cycle ranges are designed between 0.5Hz to 1000Hz, 0.1mA to 4mA, and 10% to 90% respectively. Output signals can also be patterned into one of RANDOM, F.M., or Burst types. The burst pattern is configurable for its frequency and pulses train counts. In comparison with counterpart devices in the market, our device have more options and more adjustable just on a device with the use of no PC or any kind of programmer with less price.

## ACKNOWLEDGMENT

We would like to express our great appreciation to Mr. Morteza Mojaver, for his valuable and constructive suggestions during the planning and development of this work. His willingness to give his time so generously has been very much appreciated. Assistance provided by Dr. Vahid Abootalebi, Dr. Mahdi Jafari, and Ms. Mahnoosh Zarifnasab was greatly appreciated.